\documentclass[10pt,sigconf,authorversion]{acmart}

\usepackage{subcaption}

\usepackage{xcolor}
\usepackage{framed}

\usepackage[utf8]{inputenc}
\usepackage{enumitem}
\AtBeginDocument{%
  \providecommand\BibTeX{{%
    \normalfont B\kern-0.5em{\scshape i\kern-0.25em b}\kern-0.8em\TeX}}}

\usepackage[printonlyused]{acronym}
\usepackage{tikz}
\usepackage{amsmath}
\usepackage{subcaption}
\usepackage{tabularx} 
\usepackage{tabularray}
\usepackage{multirow}
 \usepackage{colortbl} 
\usepackage{tikz}
\usepackage{array}

\newlength{\leftbarwidth}
\newlength{\leftbarsep}
\colorlet{leftbarcolor}{black}

\renewenvironment{leftbar}{%
    \MakeFramed {\advance \hsize -\width \FrameRestore }%
}{%
    \endMakeFramed
}

\begin{document}

\copyrightyear{2024}
\acmYear{2024}
\setcopyright{acmlicensed}
\acmConference[ACM MobiCom '24]{The 30th Annual International Conference on Mobile Computing and Networking}{November 18--22, 2024}{Washington D.C., DC, USA}
\acmBooktitle{The 30th Annual International Conference on Mobile Computing and Networking (ACM MobiCom '24), November 18--22, 2024, Washington D.C., DC, USA}
\acmDOI{10.1145/3636534.3690696}
\acmISBN{979-8-4007-0489-5/24/11}

%\input{acronyms}
\title{Demystifying Privacy in 5G Stand Alone Networks}

\author{Stavros Eleftherakis}
\authornote{This work was carried out during S.~Eleftherakis' internship at Telefonica Research, Barcelona.}
\email{stavros.eleftherakis@imdea.org}
\affiliation{%
  \institution{Imdea Networks Institute}
  \institution{Universidad Carlos III de Madrid }
  \city{Madrid}
  \country{Spain}
}

\author{Timothy Otim}
\email{timothy.otim@imdea.org}
\affiliation{%
  \institution{Imdea Networks Institute}
  \city{Madrid}
  \country{Spain}
}

\author{Giuseppe Santaromita}
\email{giuseppe.santaromita@imdea.org}
\affiliation{%
  \institution{Imdea Networks Institute}
  \city{Madrid}
  \country{Spain}
}

\author{Almudena Diaz Zayas}
\email{adz@uma.es}
\affiliation{%
  \institution{Universidad de Málaga}
  \city{Málaga}
  \country{Spain}
}

\author{Domenico Giustiniano}
\email{domenico.giustiniano@imdea.org}
\affiliation{%
  \institution{Imdea Networks Institute}
  \city{Madrid}
  \country{Spain}
}

\author{Nicolas Kourtellis}
\email{nicolas.kourtellis@telefonica.com}
\affiliation{%
  \institution{Telefonica Research}
  \city{Barcelona}
  \country{Spain}
}  

\renewcommand{\shortauthors}{S. Eleftherakis et. al}

\begin{abstract}
% Preserving the privacy of users continues to be a top priority in mobile cellular networks, especially given the trends of proliferation of interconnected devices and services and the denser network infrastructure.
% In fact, a lot of user privacy issues have been raised in 2G, 3G, 4G/LTE networks.
% Recognizing this general concern, 3GPP has prioritized addressing these issues in the development of 5G, implementing
% numerous modifications to enhance user privacy since 5G Release 15.
% However, it is unclear at which level these modifications have been followed by operators, especially when considering that the majority of 5G networks still operate in 5G Non Stand Alone (NSA) mode (i.e., rely on 4G Core Networks), rather than Stand Alone (SA) mode. 
% In this paper, to the best of our knowledge, we are the first to perform a head-on qualitative and experimental comparison between 5G NSA and SA in real operator networks, with respect to enhancements introduced in 5G to address 8 top pre-5G attacks.
% We are also the first to examine the S\&P levels offered by OpenAirInterface (OAI), the de facto open source software for 5G, and experimentally compare them with 5G NSA and SA real network deployments for the same attacks.
% From our analysis, we highlight two new 5G S\&P vulnerabilities that should be addressed with further studies and stricter 5G standards and discuss several key takeaways.

Ensuring user privacy remains critical in mobile networks, particularly with the rise of connected devices and denser 5G infrastructure.
Privacy concerns have persisted across 2G, 3G, and 4G/LTE networks.
Recognizing these concerns, the 3rd Generation Partnership Project (3GPP) has made privacy enhancements in 5G Release 15.
However, the extent of operator adoption remains unclear, especially as most networks operate in 5G Non Stand Alone (NSA) mode, relying on 4G Core Networks.
This study provides the first qualitative and experimental comparison between 5G NSA and Stand Alone (SA) in real operator networks, focusing on privacy enhancements addressing top eight pre-5G attacks based on recent academic literature.
Additionally, it evaluates the privacy levels of OpenAirInterface (OAI), a leading open-source software for 5G, against real network deployments for the same attacks.
The analysis reveals two new 5G privacy vulnerabilities, underscoring the need for further research and stricter standards.
\end{abstract}


\maketitle

%\vspace{-3.5mm}
\section{Introduction}

Telecommunication companies worldwide are rapidly rolling out 5G technology, with projections already suggesting billions of active 5G connections based on 3GPP Release 15~\cite{5gamericas}.
This faster, denser, and larger-bandwidth cellular technology is being used by the ever increasing number of IoT, smartphone, and other connected devices, estimated to be 15 billion by the end of 2023, and projected to reach 30 billions by 2030~\cite{statistic_IoT_devices}.
However, the pressure to fast deploy 5G networks comes at a cost.
Although operators started to deploy and use 5G networks, a lot of these are 5G Non Stand Alone (NSA), i.e., they rely on 4G Core Networks to operate.
For example, in Europe in 2022, the 5G Stand Alone (SA) deployment had reached only a $38\%$ with respect to the total number of 5G Networks, although this is increasing~\cite{5G_stats}.
In fact, 2024 is expected to be the year that 5G SA deployment will accelerate~\cite{5gsa}.

Continued reliance on 5G NSA technology poses a significant security and privacy (S\&P) challenge, stemming from unresolved adversarial issues carried over from pre-5G networks.
Despite efforts to address these S\&P issues with the introduction of 5G, transitional 5G NSA networks still exhibit vulnerabilities.
Prior research on pre-5G networks has revealed numerous S\&P concerns regarding user confidentiality, authentication, integrity, location privacy, anonymity, and user unlinkability~\cite{shaik2019new,karakoc2023never,kune2012location,arapinis2017analysis,shaik2015practical,bojic2017opportunities}.
Indeed, while 3GPP incorporated some of these findings to enhance 5G S\&P features in Release~15 (Rel.~15), the promise of improved S\&P mainly applies to 5G SA, not NSA networks.
In fact, NSA's reliance on 4G core infrastructure increases the potential attack surface.

Given the proliferation of connected devices and shift from 5G NSA to SA, several key questions motivate this study:
1) What are the core differences in S\&P aspects between 5G NSA and SA networks?
2) How have industry standards and academic literature addressed pre-5G vulnerabilities in 5G real deployments, and what methods have been employed for enhancement?
3) Do current open-source tools, readily available to the research community such as OpenAirInterface (OAI)~\cite{OAI}, adequately support and enable proper measurement capabilities for evaluating these 5G SA enhancements?
4) Are there any novel S\&P flaws in 5G SA that remain unidentified and unresolved?
With this work, we dive into the S\&P aspects of 5G SA and NSA networks in Rel.~15 and, to the best of our knowledge, make the following contributions:
\setlist{nolistsep}
\begin{itemize}
   
    \item We are the first to perform a head-on qualitative and experimental comparison between 5G NSA and SA in real operator networks, and study enhancements introduced in 5G to address 8 top pre-5G attacks as described in recent literature~\cite{eleftherakis2024sokevaluating5gprotocols,khan2020survey,khan2018identity,rupprecht2018security,yu2021improving,behrad2018securing}. %For this, we propose a methodological framework, and describe the experimental testbeds used.
    \item We are the first to perform a study on the security aspects of real 5G SA networks in Europe and the second in the world.
    \item~ We are the first to show the implementation of SUCI in a real 5G network.
    \item We are the first to examine the S\&P levels offered by OpenAirInterface (OAI), the de facto open source software for 5G, and experimentally compare them with 5G NSA and SA real deployments for the same attacks.
    \item We highlight two new 5G S\&P vulnerabilities that merit further research and stricter 5G standards.
    \item We discuss 9 key takeaways from our overall analysis.
\end{itemize}

%\vspace{-6mm}
\section{Background}

\begin{comment}

\begin{figure}
  \centering
  \includegraphics[width=0.9\linewidth]{pictures/paper_network.pdf} 
  \caption{Overview diagram of 5G cellular network infrastructure.}
  \label{fig:network_representation}
\end{figure} 
\end{comment}

%Here, we cover necessary concepts for the rest of the paper. 

\subsection{5G Cellular Network Architecture}

We provide an overview of the main infrastructure of a 5G Cellular Network: User Equipment (UE), New Generation Radio Access Network (NG-RAN), and Core Network (CN).\\ %and their properties, as depicted in Figure~\ref{fig:network_representation}:\\
\noindent
\textbf{User Equipment (UE):} The UE consists of the Mobile Equipment (ME) and the Universal Subscriber Identity Module (USIM) card. It is used by consumers to access mobile services and applications. The USIM card pays a pivotal role in authentication, key generation and subscriber information management. 
%For the purpose of this paper, we consider that the ME is a smartphone.
%The ME encompasses both hardware components (e.g.,~processor, memory) and software components (e.g.,~operating system, applications).

\noindent
\textbf{New Generation Radio Access Network (NG-RAN):} The New Generation Radio Access Network (NG-RAN) encompasses a network of Base Stations (BSs), known as gNBs in 5G, that are organized into distinct Tracking Areas (TAs).  The RRC (Radio Resource Control) protocol~\cite{3GPP_RRC} plays a critical role in establishing, maintaining, and terminating radio connections between UEs and the NG-RAN.

%Its responsibilities include resource management, facilitating seamless handovers, provisioning wireless connectivity, and enabling communication between UEs and the Core Network (CN). A key component of the NG-RAN is the BSs i.e., gNBs in 5G.
%Within the NG-RAN, various components such as base stations (e.g., gNBs in 5G), antennas, and a multitude of hardware and software elements enable radio communication. The RRC (Radio Resource Control) protocol~\cite{3GPP_RRC} plays a critical role in establishing, maintaining, and terminating radio connections between UEs and the NG-RAN.

\noindent
\textbf{Core Network (CN):} The CN comprises diverse entities known as Network Functions (NFs), each tasked with delivering different network services and functionalities such as mobility management, authentication, subscriber data management, session establishment and control, etc. The Non-Access Stratum (NAS) protocol serves as the pathway facilitating signaling procedures and the exchange of messages between the UE and the CN and a comprehensive analysis of NAS protocol can be found in~\cite{3GPP_NAS}. %In the next section we focus on NFs that are related to privacy. 

\vspace{-4mm}
\subsection{UE Identifiers}\label{Sec:Identifiers}

There are various identifiers used in the Cellular Network, responsible for the identification of different entities participating in the system, and especially of the UE at hand. In general, they are divided into permanent and temporary identifiers. 
Permanent identifiers are global and are considered as extremely sensitive in terms of privacy. Temporary identifiers are used in order to minimize the transmission of permanent ones, thus enhancing UE privacy, but certain rules should be followed for them as well.

To begin with, 5G introduced Subscription Unique Permanent Identifier (SUPI) in the clause 5.9.2 of the 3GPP 5G technical specifications for security architecture and procedures~\cite{5GS_architecture}. SUPI is the permanent identity of the USIM card and must never be submitted plaintext, except for emergency cases, as denoted in the clause 5.2.5 of~\cite{3gpp_fake_base_stations}. SUPI should not be used for the authentication of the UE.

Another important permanent identifier is the Permanent Equipment Identity (PEI), manufactured on the ME device during its production as analyzed in Sec.~6.4 of~\cite{Numbering}. PEI should only be transmitted in a secure channel, after integrity and ciphering have been enabled, and cannot be used for the authentication of the UE by the network. Intuitively, SUPI and PEI are equivalent to the International Mobile Subscriber Identity (IMSI) and International Mobile Equipment Identity (IMEI), respectively in previous generations.

The second important category of identifiers are temporary. First, 5G introduces the Subscription Unique Concealed Identifier (SUCI) in the clause 5.9.2a of~\cite{5GS_architecture}. SUCI is an elliptic, cryptography-based concealed version of SUPI that is constructed by the USIM card and is used for the authentication of the UE by the network. The concept of SUCI is new in 5G and nothing equivalent exist in previous generations. 

Furthermore, the CN assigns a temporary identifier to the UE for the communication between the UE and the CN, called 5G-GUTI (5G Globally Unique Temporary Identifier) as defined in the clause 5.9.4 of~\cite{5GS_architecture}. In 2G and 3G, 5G-GUTI was called Temporary Mobile Subscriber Identity (TMSI), whereas in 4G it was called GUTI or TMSI. For the rest of this paper, we denote this temporary identifier as TMSI for previous to 5G cellular generations. Finally, the UE is also assigned a temporary identifier called Cell Radio Network Temporary Identity (C-RNTI) from the RAN, facilitating the communication between the UE and the RAN.  

\vspace{-3mm}
\subsection{UE Capabilities}\label{Sec:Device_Capab}

The capabilities of an UE can be separated into network capabilities~\cite{3GPP_NAS} and radio access capabilities~\cite{3GPP_Radio_Access_Capab}.
The network capabilities indicate general UE characteristics, such as the security algorithms supported by the UE for integrity and ciphering protection, and are transmitted as a NAS message. 

As for the security algorithms, 5G UEs shall support New Radio Encryption Algorithm (NEA) 0, 128-NEA1 and 128-NEA2 for ciphering (confidentiality) protection and New Radio Integrity Algorithm (NIA) 0, 128-NIA1 and 128-NIA2 for integrity protection (Secs.~5.2.2 and~5.2.3 of~\cite{3gpp_fake_base_stations}).

The radio access capabilities contain information regarding the radio capabilities of the UE, such as the supported frequency bands of the UE, and are transmitted as an RRC message. As explained next, the UE reports its capabilities to the network during the registration procedure. Further information about the UE capabilities can be found in~\cite{shaik2015practical,shaik2019new}.

\begin{figure}
  \centering
  \includegraphics[width=1.0\linewidth]{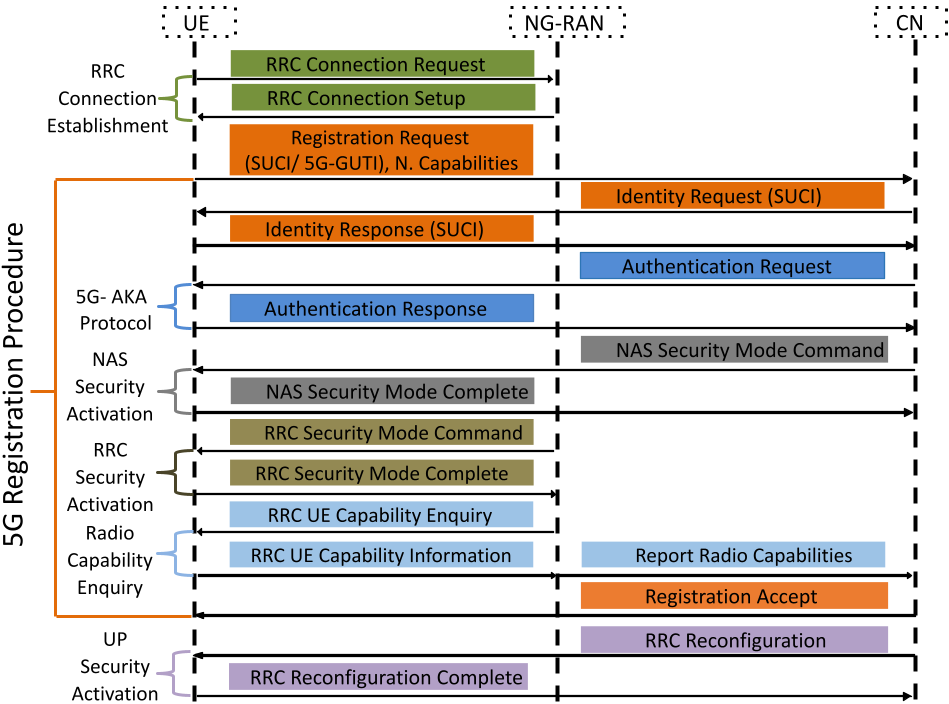} 
  \caption{High Level Representation of 5G messages exchange flow.}
  \label{fig:5G_flow}
  \vspace{-6.5mm}
\end{figure} 

\vspace{-3mm}
\subsection{5G Message Exchange Flow}\label{Sec:message_flow}

This section describes some basic message exchange flow regarding some critical procedures in 5G, as introduced in the 3GPP 5G specifications~\cite{3gpp_fake_base_stations} and illustrated in Figure~\ref{fig:5G_flow}.
First, the UE establishes an RRC Connection with the NG-RAN, and afterwards transmits a registration request message including a subscriber identity (SUCI or 5G-GUTI) and its network capabilities. % as described in Sec.~\ref{Sec:Device_Capab}.
 If the 5G-GUTI was transmitted by the UE and the CN cannot resolve it, it sends to the UE an Identity Request message. 
The UE transmits the SUCI in a Identity Response message, and then the Authentication and Key Agreement Protocol (AKA) is initiated by the CN. 
We highlight that both EPS-AKA (Evolved Packet System Authentication and Key Agreement) and 5G-AKA (5G Authentication and Key Agreement) can be used, but since the differences are out of scope for this paper, 5G-AKA (Sec.~6.1.3.2 of~\cite{3gpp_fake_base_stations}) is referred. Further information about the 5G-AKA protocol can be found in~\cite{wang2021privacy,basin2018formal,koutsos20195g}.
%In a nutsell, the network sends a random number (RAND) and an authentication token (AUTN) to the UE. The UE first verifies the validity of RAND and the freshness of AUTN and if the verification is not successful, it sends a MAC failure or Sync Failure respectively. If both of them are verified successfully, the UE uses the RAND and its permanent key to generate a response (RES) and sends it to the network as its Authentication Response.The network verifies the validity of RES and, if it is valid, the authentication is successful. 

After a successful authentication response sent by the UE, NAS Security Mode Command (SMC) is transmitted in order to initiate the activation of a secure channel for the NAS protocol messages, providing integrity and ciphering protection. As analyzed in the Sec.~6.7.2 of~\cite{3gpp_fake_base_stations} an enhanced NAS SMC procedure is followed, where the CN replays the Network capabilities received in the first Registration message to ensure that they were not modified. Furthermore, a Message Authentication Code (MAC) is included to ensure the integrity of the NAS SMC message. The UE verifies the correctness of the Network capabilities and the integrity of the NAS SMC message and if everything is fine, a secure NAS channel is established. 

The same procedure is followed for the RRC messages, exchanged between the UE and the NG-RAN, for the activation of a secure channel for them as well. Then, the UE radio capabilities are transmitted to the 5G network, \textit{after the establishment of a secure RRC channel} (see Figure~\ref{fig:5G_flow}). This is in contrast to prior cellular generations, where radio capabilities were sent before the establishment of a secure channel between the UE and the RAN. Finally, integrity and ciphering for the User Plane (UP) messages are activated through the RRC Reconfiguration message.

\vspace{-3mm}
\subsection{Paging Procedure}

The paging mechanism notifies the receiving UE for incoming data transmissions or phone calls. These paging messages are primarily transmitted by the base station (RRC paging), broadcasting the UE's identity and indicating the recipient of incoming data or a call via the paging channel. 
After data or an SMS are sent to a UE, the paging procedure is activated at the incoming or receiving UE, prompting all UEs within the cell to monitor the paging channel and respond if their identity matches.
Similarly for phone calls, this process occurs at a TA level. As discussed later, the paging mechanism posed significant privacy concerns across various cellular generations, as it relied on IMSI, potentially enabling adversaries to compromise the UE's permanent identity.

%After receiving incoming data or an SMS, the paging procedure is activated, prompting all UEs within the cell to monitor the paging channel and respond if their identity matches.

\begin{figure*}
  \centering
  \includegraphics[width=0.99\linewidth]{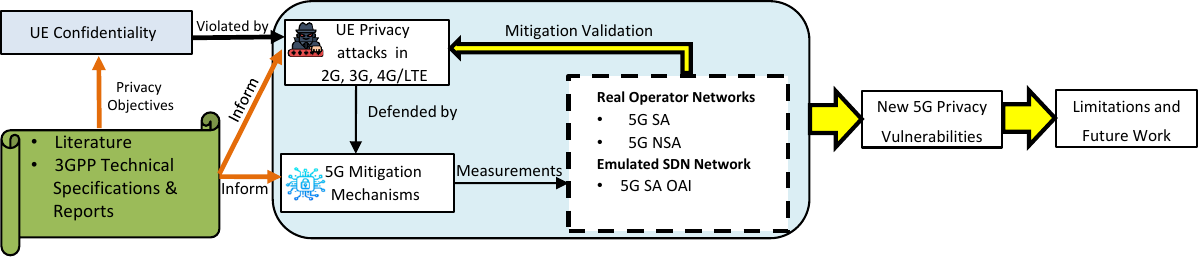} 
  \caption{Framework followed to study pre-5G attacks in real 5G NSA \& SA deployments, and an OAI testbed.}
  \vspace{-3mm}
  \label{fig:5G_framework}
\end{figure*} 

%\vspace{-3mm}
\section{Methodology}

In this section, we describe the methodology of our work. 
First, in Sec.~\ref{Sec:privacy_objectives}, we define User Identity Confidentiality and its properties.
Then, Sec.~\ref{Sec:Framework} provides an end-to-end overview of the Framework we use, thereby navigating the reader for the rest of this paper.
Finally, Sec.~\ref{Sec:Testbeds} describes the different 5G SA and NSA deployment scenarios, and the common experimental process used in various experiments executed.

\subsection{User Identity Confidentiality}\label{Sec:privacy_objectives}

%Given the multitude of interconnected devices present in modern networks and the extensive array of sensitive data being transmitted,
It is crucial to establish certain S\&P objectives about the confidentiality of the user identity in mobile networks.
Thus, Sec.~5.1.1 of~\cite{3g_security} describes some important security characteristics that should be met by cellular networks.
In fact, as previous works~\cite{eleftherakis2024sokevaluating5gprotocols,khan2020survey,khan2018identity,borgaonkar2018new} considered them, the following properties are necessary for the protection of the user identity confidentiality:
\begin{enumerate}
    \item UE Identity Privacy: "the permanent user identity (IMSI) of a user to whom a services is delivered cannot be eavesdropped on the radio access link".
    \item UE Location Privacy: "the presence or the arrival of a user in a certain area cannot be determined by eavesdropping on the radio access link".
    \item UE Untraceability: "an intruder cannot deduce whether different services are delivered to the same user by eavesdropping on the radio access link".
\end{enumerate}

%\vspace{-3mm}
\subsection{Framework}\label{Sec:Framework}

The framework followed in this paper is outlined in Figure~\ref{fig:5G_framework}:
First, we search in related surveys~\cite{eleftherakis2024sokevaluating5gprotocols,rupprecht2018security,khan2019survey,khan2020survey} for pre-5G related attacks that violated UE Confidentiality.
Then, in Sec.~\ref{Sec:pre-5g-attacks}, we identify the corresponding 5G Security mechanisms that have been planted in 5G to mitigate these attacks, again by looking into related literature (e.g.,~\cite{nie2022measuring,lasierra2023european}) and 5G 3GPP Technical Specifications (e.g.,~\cite{3gpp_fake_base_stations}). 
Afterwards, we perform measurements in five different deployment scenarios (outlined in Sec.~\ref{Sec:Testbeds} below). %and collect appropriate data. % as described in Sec.\ref{Sec:Collection_Process}. 
In Sec.~\ref{Sec:pre-5g-attacks}, we make a qualitative analysis of these measurements examining if the mitigation mechanisms embedded in 5G are implemented correctly, thus mitigating the above-mentioned attacks, inherited from past generations of cellular networks. Building on this knowledge, Sec.~\ref{sec:new_vulnerabilities} outlines two new privacy vulnerabilities that we identified through the measurements in the real operator networks. Finally, Sec.~\ref{sec:discussion} discusses lessons learned and directions for future research work, regulation and standardization. We remark that the proposed framework is the first that compares 5G SA and 5G NSA resilience against well-known pre-5G attacks in the same operator's networks. It is also the first framework that identified novel 5G privacy attacks in 5G SA real operator networks and the first that analyzed OAI privacy compliance.

We mention that our framework can be used for analyzing the existence of the specific pre-5G (Sec. 4) or new 5G attacks (Sec. 5) in different operators’ networks as well. Indeed, the specifics
of which messages and actions need to be checked for each attack are further analyzed experimentally
in the main body of Secs. 4 and 5, and can be repeated by other researchers/operators. Finally, the
following Sec. 3.3.3 (Data Collection Approach) attempts to homogenize the experimental process in
order to facilitate in higher degree the reproduction of our framework.

\subsection{Experimental \& Deployment Scenarios}\label{Sec:Testbeds}

Next, we describe the different 5G testbeds that were used for our experiments. Further, we explain the data collection approach during experiments.

\subsubsection{Operator's networks testbed:} Real private 5G SA and NSA networks from a large operator in a city of Spain. We test three 5G SA networks and one 5G NSA. All networks are based on 5G Rel.~15, operate in a single TA and only data transmission is supported. We use an Asus exp21 5G smartphone equipped with a 5G USIM card to connect to these networks. The smartphone is equipped with the Nemo Handy diagnostic tool~\cite{NemoHandy}, that allows measuring information on wireless diagnostics of air interface and mobile application quality-of-service and quality-of-experience.

\subsubsection{5G SA in OpenAirInterface (OAI):} A SDN testbed based on OAI in a lab setting. We connect a Google Pixel 7 with a 5G USIM card. Our analysis is based on CN and gNB log files and wireshark captures. OAI has been chosen for this work because it has become the de facto open-source software solution for 5G applications. To build the SA network, the gNB consisted of an Ettus N310 SDR and a high-performance Intel Core i7-8700k 3.7 GHz CPU work station. To establish the 5G CN, an Intel Core i7-8700k 3.7 GHz CPU workstation with Linux operating systems serving as the host as used. This workstation emulated all the required functionalities of a 5G NR CN. When the smartphone connected to the OAI 5G network, the CN and gNB log files were saved for post processing analysis. 

\subsubsection{Data Collection Approach}\label{Sec:Collection_Process}

To facilitate and homogenize the data collection procedure, the following experimental process has been followed across experiments:
\begin{itemize}
      \item Step 1: Airplane mode ON: The terminal will always start with airplane mode activated.
     \item Step 2: Start data collection: Once the airplane mode of the UE is deactivated, the data collection starts.
     \item Step 3: Time duration of the experiment: Based on the nature of the experiment, we wait from some seconds to some days.
     \item Step 4: Airplane mode ON: End of experiment and stop of data collection process.
\end{itemize}

\begin{comment}
    
\subsection{Adversaries}

\begin{figure}[t]
\begin{subfigure}{0.5\textwidth}
  \centering
  \includegraphics[width=.8\linewidth]{pictures/Passive adversary.pdf}  
  \caption{Passive adversary.}
  \label{fig:passive_adversary}
\end{subfigure}
\hfill
\begin{subfigure}{0.5\textwidth}
  \centering
  \includegraphics[width=.8\linewidth]{pictures/Semi-Passive adversary.pdf}  
  \caption{Semi-passive adversary.}
  \label{fig:Semi_passive_adversary}
\end{subfigure}
\hfill
\begin{subfigure}{0.5\textwidth}
  \centering
  \includegraphics[width=0.8\columnwidth]{pictures/Man in the middle.pdf}  
  \caption{Active adversary.}
  \label{fig:active_adversary}
\end{subfigure}
\caption{Different types of adversaries in wireless networks.}
\label{fig:adversaries}
\end{figure}
\end{comment}

% \usepackage{multirow}
\begin{table*}[]
\begin{tabular}{|c|cc|cc|c|}
\hline
\multirow{2}{*}{\textbf{Attack Name}}                                                                               & \multicolumn{2}{c|}{\textbf{5G Security}}                                                                                                                                                               & \multicolumn{2}{c|}{\begin{tabular}[c]{@{}c@{}}\textbf{Operator's Implementations}\\ \textbf{Supported Features}\end{tabular}} & \multicolumn{1}{c|}{\begin{tabular}[c]{@{}c@{}}\textbf{Emulated}\\ \textbf{SDN Testbed}\end{tabular}} \\ \cline{2-6} 
                                                                                                           & \multicolumn{1}{c|}{\textbf{Mitigation Mechanisms}}                                                              & \multicolumn{1}{c|}{\begin{tabular}[c]{@{}l@{}}\textbf{Optional or}\\ \textbf{Mandatory}\end{tabular}} & \multicolumn{1}{c|}{\textbf{5G NSA}}                                      & \textbf{5G SA*}                                     & \textbf{OAI 5G SA}                                                                           \\ \hline \hline
\begin{tabular}[c]{@{}c@{}}IMSI Catching\\ {~\cite{strobel2007imsi,mitchell2001security,patent,paget2010practical,meyer2004man,kotuliak2022ltrack,erni2022adaptover,hussain2019privacy}}\end{tabular}                               & \multicolumn{1}{c|}{SUCI}                                                                               & O                                                                                    & \multicolumn{1}{l|}{No}                                          & Yes                                       & Yes                                                                                 \\ \hline
\begin{tabular}[c]{@{}c@{}}IMSI Paging\\ ~\cite{kune2012location,arapinis2017analysis,shaik2015practical,bojic2017opportunities}\end{tabular}                                 & \multicolumn{1}{c|}{\begin{tabular}[c]{@{}c@{}}5G-TMSI based \\ paging\end{tabular}}                    & M                                                                                    & \multicolumn{1}{l|}{Yes}                                         & Yes                                       & Yes                                                                                 \\ \hline
\begin{tabular}[c]{@{}c@{}}IMEI Catching\\ \cite{dabrowski2014imsi,olimid2017lowcost,michau2016not,park2022doltest}\end{tabular}                               & \multicolumn{1}{c|}{IMEI in secure channel}                                                             & M                                                                                    & \multicolumn{1}{l|}{Yes}                                         & Yes                                       & Yes                                                                                 \\ \hline
\begin{tabular}[c]{@{}c@{}}TMSI Deanonymity\\ \cite{kune2012location,arapinis2012new,arapinis2014privacy,hong2018guti}\end{tabular}                            & \multicolumn{1}{c|}{\begin{tabular}[c]{@{}c@{}}GUTI reallocation \\ mechanism\end{tabular}}             & M                                                                                    & \multicolumn{1}{l|}{No}                                          & Yes                                       & No                                                                                  \\ \hline
\begin{tabular}[c]{@{}c@{}}C-RNTI Tracking\\ \cite{jover2016lte_first,jover2016lte,rupprecht2019breaking,rupprecht2020call}\end{tabular}                             & \multicolumn{1}{c|}{\begin{tabular}[c]{@{}c@{}}RRC Ciphering\end{tabular}}                     & O                                                                                    & \multicolumn{1}{l|}{No}                                          & No                                        & No                                                                                  \\ \hline
\begin{tabular}[c]{@{}c@{}}UE Measurements reports\\ \cite{shaik2015practical,olimid2017lowcost,bitsikas2021don}\end{tabular}                     & \multicolumn{1}{c|}{RRC Ciphering}                                                                      & O                                                                                    & \multicolumn{1}{l|}{No}                                          & No                                        & No                                                                                  \\ \hline
\begin{tabular}[c]{@{}c@{}}Security capabilities\\ bidding down attack\\ ~\cite{shaik2015practical,shaik2019new,karakoc2023never}\end{tabular} & \multicolumn{1}{c|}{Enhanced NAS process}                                                               & M                                                                                    & \multicolumn{1}{l|}{Weak}                                        & Weak                                      & Yes                                                                                 \\ \hline
\begin{tabular}[c]{@{}c@{}}Radio Capabilities bidding\\ down attack\\ ~\cite{shaik2019new,karakoc2023never}\end{tabular}    & \multicolumn{1}{c|}{\begin{tabular}[c]{@{}c@{}}Radio Capabilities \\ in a Secured Channel\end{tabular}} & M                                                                                    & \multicolumn{1}{l|}{Yes}                                         & Yes                                       & Yes                                                                                 \\ \hline
\end{tabular}

\begin{flushleft}
$\ast$: tested over three different 5G SA networks. The best result achieved among them is reported.
\end{flushleft}
\caption{Summary of pre-5G attacks, 5G mitigation strategies, and employment in different testbeds and networks.}
\vspace{-8mm}
\label{tab:Overview}
\end{table*}

\section{Pre-5G Attacks \& 5G Enhancement}\label{Sec:pre-5g-attacks}

We focus on 8 top adversarial attacks in pre-5G cellular networks, that have been documented extensively in academic literature~\cite{eleftherakis2024sokevaluating5gprotocols,khan2020survey,khan2018identity,rupprecht2018security,yu2021improving,behrad2018securing}.
These attacks attempt to break either the whole set or a portion of the UE Confidentiality properties described in Sec.~\ref{Sec:privacy_objectives}.
Here, and for each attack, we first outline the attack and then refer to the corresponding 5G mitigation mechanisms as analyzed in~\cite{eleftherakis2024sokevaluating5gprotocols,khan2020survey,rupprecht2018security,yu2021improving}.
Then, we examine the existence and correct implementation of each specific 5G mitigation mechanisms across the different deployment scenarios described earlier in Sec.~\ref{Sec:Testbeds}. In Table~\ref{tab:Overview}, we summarize our findings for all attacks.

\subsection{Attacks on Permanent Identifiers}
\subsubsection{\textbf{IMSI Catching:}} \label{Sec:IMSI_Catchers}

In this section, we refer to attacks that aim to steal the IMSI of the UE, known as IMSI catching attacks~\cite{mitchell2001security,patent,lilly2017imsi,park2019anatomy}. 
Indeed, the attacker uses a device called IMSI Catcher consisting of a fake base station, thus being easily deployable and affordable~\cite{dabrowski2014imsi,paget2010practical,olimid2017lowcost}.
A fake base station can be constructed by using a Universal Software Radio Peripheral (USRP)~\cite{USRP} with a modified code of open-source projects like OpenLTE~\cite{OpenLTE}, srsRAN~\cite{srsRAN,gomez2016srslte}, gr-LTE~\cite{gr-LTE,demel2015lte}, or OAI~\cite{OAI,nikaein2014openairinterface}.
In fact, IMSI Catching~\cite{mitchell2001security,patent,lilly2017imsi} has been a persistent attack in 2G~\cite{meyer2004man}, 3G~\cite{ahmadian2009new} and 4G~\cite{michau2016not}.
The adversary has two different ways to steal the IMSI of the UE.
First, and more usual, an IMSI catcher takes advantage of the victim's phone behavior to connect to the Cell that offers the strongest signal power. When the UE connects to the IMSI catcher device, the adversary sends an Identity Request message. Then, the UE answers with an Identity Response message, including the IMSI without encryption (plaintext), thus, leading to UE identity disclosure. 

As an alternative, signal overshadowing techniques have been used as well~\cite{yang2019hiding}. This kind of attacks require time and frequency synchronization with the legitimate Base Station (BS), offering signal strength slightly stronger~\cite{yang2019hiding} or slightly weaker~\cite{ludant2021sigunder} than the legitimate one. 
In fact, IMSI catching based on signal overshadowing~\cite{erni2022adaptover,kotuliak2022ltrack} is stealthier compared to the traditional, fake-BS-based attack, since it uses a normal signal strength, thus making its detection even more difficult. 

A lot of different solutions to IMSI catchers had been proposed in the literature for IMSI Catcher detection, but all of them suffer from practicality problems. A first set of solutions proposed either the usage of multiple IMSIs~\cite{khan2015improving,multipleIMSIs} or the introduction of a new pseudonym instead of the IMSI~\cite{van2015defeating,norrman2016protecting}, but both of them suffer from synchronization problems between the USIM and the network as described in~\cite{khan2017trashing,hussain2018lteinspector}. Furthermore, other solutions
~\cite{ekene2016enhanced,zhang2005security,deng2009novel,choudhury2012enhancing,geir2013privacy,li2011security} proposed significant changes to the AKA protocol, thus making their implementation impractical. Finally, Dabrowski et all.~\cite{dabrowski2014imsi,dabrowski2016messenger} proposed an IMSI Catcher detection framework based on possible network abnormalities (e.g.,~strange Cell frequencies, unusual Cell locations and Cell IDs, signal noise level, unusual network parameters) that could have been created by the existence of an IMSI Catcher. It is unclear if such a detection framework has been implemented by any operator. Based on this outlook, the plaintext IMSI transmission was characterized as a key vulnerability in the clause 6.1.3 of 3GPP Specifications~\cite{3gpp_IMSI_catching_problem}. 

\setlength{\leftbarwidth}{5pt}
\setlength{\leftbarsep}{8pt}
\colorlet{leftbarcolor}{black}

\begin{leftbar}
\noindent \textbf{5G Enhancement 1: SUPI Concealment} \\ 
SUPI is the corresponding Identifier to IMSI in 5G networks. In order to avoid the plaintext transmission of SUPI, Subscriber Unique Concealed Identifier (SUCI) is introduced as an~\textit{encrypted} form of SUPI, based on elliptic cryptography. In fact, SUCI can be mainly used for authentication if the temporary identifier, 5G-GUTI, is not available. SUCI is an optional feature based on 3GPP TSs. 
\end{leftbar}

\begin{figure}
  \begin{subfigure}{0.45\textwidth}
  \includegraphics[width=0.99\linewidth]{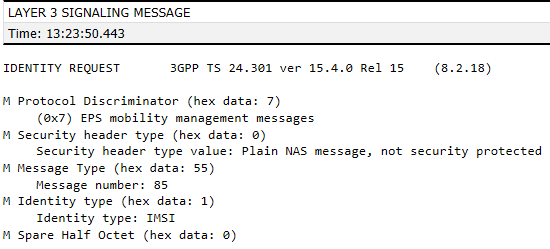} 
  \caption{IMSI based Identity Request in 5G NSA operator's network.}\label{fig:5G_NSA_IMSI}
  \end{subfigure}
  \begin{subfigure}{0.45\textwidth}
  \includegraphics[width=0.99\linewidth]{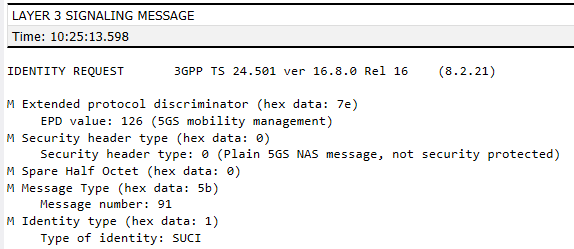}
  \caption{SUCI based Identity Request in 5G SA operator's networks.}\label{fig:5G_SA_SUCI}
  \end{subfigure}
  \begin{subfigure}{0.45\textwidth}
  \includegraphics[width=0.99\linewidth]{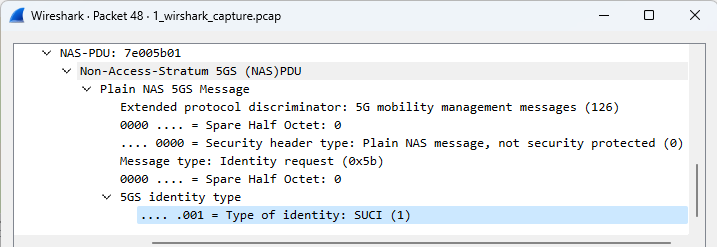}%\label{fig:5G_OAI_SUCI} 
  \caption{SUCI based Identity Request in OAI based SA network.}\label{fig:5G_OAI_SUCI} 
  \end{subfigure}
  \caption{Identity Request in 5G NSA, SA and OAI networks.}\label{fig:Identity_Request}
  \vspace{-5mm}
\end{figure}

\noindent \textbf{Experimental Validation:}
We performed multiple UE registrations in all the different deployment scenarios available and analyzed the Identity Request message content, as depicted in Fig.~\ref{fig:Identity_Request}.
First, we find that the 5G NSA operator's implementation does not support SUCI, transmitting the Identity Request with IMSI as illustrated in Fig.~\ref{fig:5G_NSA_IMSI}, and thus being vulnerable to IMSI catchers attacks. Then, as illustrated in Fig.~\ref{fig:5G_SA_SUCI} and Fig.~\ref{fig:5G_OAI_SUCI}, all operator's 5G SA implementations and OAI support SUCI, thereby eliminating the privacy risks imposed by IMSI Catchers. To the best of our knowledge, ours is the first study showing that an operators' network supports this important optional feature, since a previous study among Chinese 5G SA networks showed that SUCI was not supported~\cite{nie2022measuring}.
 
Existing literature has emulated some attacks against SUCI.
For instance, a SUCI probing or SUCI replay attack~\cite{chlosta20215g} tried to obtain the victim's SUCI and verify if a Person of Interest (PoI) is in a current location or not. 
3GPP characterized this attack (Key Issue \#2.2 of~\cite{imsiProbing3GPP}) as low risk and no normative measures are needed. The reason behind this is that obtaining the actual identity of the user cannot be revealed by the SUCI, thus, there is no threat of UE identification.
Furthermore, SUPI guessing attacks have been discussed in the literature as well~\cite{khan2018identity,liu2021security}.
The adversaries first guess a SUPI and generate SUCIs from this. 
Then, the produced SUCIs are sent to potential victims, trying to verify if the guessed SUPI belongs to the victim user.
Such attacks have been analyzed as a Key Issue \#3.2 in~\cite{imsiProbing3GPP}, concluding that no normative measures should be taken against them since the likelihood of their accuracy is small.

\setlength{\leftbarwidth}{5pt}
\setlength{\leftbarsep}{8pt}
\colorlet{leftbarcolor}{blue}

\begin{leftbar}
\noindent \textbf{Takeaway 1:}
The SUCI mechanism significantly improves the identity privacy of UEs; it is crucial that real operator CNs support this privacy-enhancing feature.
\end{leftbar}

\vspace{-4mm}
\subsubsection{\textbf{Attacks based on IMSI Paging:}}\label{Sec:IMSI_paging_attack}

Paging procedure is initiated when the network searches for an UE in order to deliver a service to the device, such as a phone call or an SMS.
In general, the temporary identifier (TMSI) is used for paging, but in previous generations to 5G, there are cases where (e.g.,~TMSI cannot be resolved by the network) IMSI can be used as well.
The fact that IMSI could be sent in cleartext in paging messages made the paging process vulnerable, as were shown in 2G~\cite{kune2012location}, 3G~\cite{arapinis2012new,arapinis2017analysis} and 4G~\cite{shaik2015practical,bojic2017opportunities}.
The attacker initiates the paging process by sending messages, or making phone calls to the victim and at the same time a sniffer~\cite{ludant20235g,hoang2023ltesniffer,kumar2014lte,bui2016owl,falkenberg2019falcon} can observe the unencrypted downlink paging messages and identify the IMSI of the victim's UE. %More information about silent calls and messages can be found in~\cite{hong2018guti}, but in summary, it is a call or message that activates the paging mechanism without the recipient to get notified.

\setlength{\leftbarwidth}{5pt}
\setlength{\leftbarsep}{8pt}
\colorlet{leftbarcolor}{black}

\begin{leftbar}
\noindent
\textbf{5G Enhancement 2: Decoupling IMSI from paging}
\\The above-mentioned problem was taken into consideration in 5G and  the decoupling of the IMSI/SUPI from the paging mechanism is proposed. Indeed, in 5G paging takes place with a shortened version of 5G-GUTI, called 5G-S-TMSI (5G S-Temporary Mobile Subscription Identifier) as mentioned in Sec.~2.10.1 of~\cite{Numbering}. 5G-S-TMSI is derived from 5G-GUTI, so its strict update mechanism as analyzed in the Section~\ref{5G_GUTI_update} holds for 5G-S-TMSI as well.
\end{leftbar}

\noindent \textbf{Experimental Validation:}
To verify this functionality, we sent messages to the victim's phone, while it was connected to each network, through the Messenger application~\cite{MessengerApp}.
All of our networks used 5G-S-TMSI for the paging message transmission. After many different experiments, IMSI or SUPI were never used for paging.

\setlength{\leftbarwidth}{5pt}
\setlength{\leftbarsep}{8pt}
\colorlet{leftbarcolor}{blue}

\begin{leftbar}
\noindent \textbf{Takeaway 2:}
Both 5G NSA and SA real networks, and the OAI open-source implementation properly follow the related 5G specification to decouple IMSI from paging~\cite{Numbering}.
\end{leftbar}

\subsubsection{\textbf{IMEI Catching:}}\label{IMEI_Catching}

IMEI (or Permanent Equipment Identity (PEI) in 5G) is another sensitive permanent identifier, corresponding to the Mobile Equipment (ME). 
In previous mobile generations, the cleartext transmission of this identifier was permitted as a response to an Identity Request Message. Thus, an active adversary using a fake base station, similar to IMSI catchers' adversaries, could send an Identity Request using the IMEI instead of the IMSI, and steal the IMEI of the ME~\cite{dabrowski2016messenger,dabrowski2014imsi}.

\setlength{\leftbarwidth}{5pt}
\setlength{\leftbarsep}{8pt}
\colorlet{leftbarcolor}{black}

\begin{leftbar}
\noindent
\textbf{5G Enhancement 3: Secrecy of PEI}\\
PEI is the corresponding identity to IMEI, that was used in previous generations. As denoted in the clause 5.2.5 of~\cite{3gpp_fake_base_stations}: ``the UE shall only send the PEI in the NAS protocol, after NAS security context is established, unless during emergency registration when no NAS security context can be established''. Therefore, the PEI should not be used in authentication in a normal scenario. 
\end{leftbar}

\noindent \textbf{Experimental Validation:}
As shown in Fig.~\ref{fig:Identity_Request}, SUCI or IMSI are used for UE authentication. PEI is requested by the network in the Security Mode Command and transmitted by UE to the CN in the Security Mode Complete message.

\setlength{\leftbarwidth}{5pt}
\setlength{\leftbarsep}{8pt}
\colorlet{leftbarcolor}{blue}

\begin{leftbar}
\noindent \textbf{Takeaway 3:}
All operator's networks and OAI open-source implementation properly follow the appropriate steps in terms of complying with the 3GPP protocols and procedures for PEI as described in~\cite{3gpp_fake_base_stations}.
\end{leftbar}
 
\subsection{TMSI Deanonymity attack}\label{Sec:TMSI_attack}

The main reason for using the temporary identifier (TMSI) is the minimization of permanent identifiers transmission, offering better anonymity to the UE. In theory, TMSI has to be periodically updated by the network to avoid UE be easily tracked and identified~\cite{arapinis2014privacy}.
However, as was shown in~\cite{arapinis2012new,arapinis2014privacy}, in some cases the TMSI remained constant even for three days, in 2G~\cite{kune2012location} and 3G~\cite{hong2018guti} during experiments that took place in different European countries.
Similar results were obtained in LTE networks~\cite{shaik2015practical,hong2018guti} as well. 
An adversary, consisting of a passive sniffer and a phone that sends (silent) calls or messages to the victim's UE, leads to linkability of the victim's phone number with its TMSI, and consequently location tracking. %As a consequence, location tracking and traceability is achieved in 2G~\cite{kune2012location}, 3G~\cite{hong2018guti} and LTE~\cite{hong2018guti,shaik2015practical}.

\setlength{\leftbarwidth}{5pt}
\setlength{\leftbarsep}{8pt}
\colorlet{leftbarcolor}{black}

\begin{leftbar}
\noindent
\textbf{5G Enhancement 4: Strict 5G-GUTI update method\\}\label{5G_GUTI_update}
5G-GUTI is the corresponding identifier to TMSI in previous cellular network generations. As analyzed before, previous generations faced serious privacy problems due to the infrequent or miss-configured refreshment of this temporary identifier. 
Based on this outlook, 5G strictly defines when the 5G-GUTI should be updated or refreshed by the Core Network in the clause 6.12.3 of~\cite{3gpp_fake_base_stations}:
\begin{itemize}
\setlength{\itemsep}{-2pt}
    \item Upon receiving Registration Request message of type ``initial registration'' or ``mobility registration update'' from a UE.
    \item Upon receiving Service Request message sent by the UE in response to a Paging message.
    \item Upon receiving Registration Request message of type ``periodic registration update'' from a UE.
    \item Upon receiving an indication from the lower layers that the RRC connection has been resumed for a UE in 5GMM IDLE mode with suspend indication in response to a Paging message.
    \item Even more frequently, based on the implementation of the operator.
\end{itemize}
In addition, it is denoted in the same 3GPP document that 5G-GUTI should be generated in an \textit{unpredictable} way.
\end{leftbar}

\noindent \textbf{Experimental Validation:}
%As for the operator's deployment scenarios, important differences were observed between 5G NSA and SA. First, as for the initial registration, both of the 5G SA networks and the 5G NSA network as well, assign a new unpredictable value of 5G-GUTI. Afterwards, we observed a significant difference between the three networks in terms of 5G-GUTI update policy during paging. The 5G NSA implementation never assigned a new 5G-GUTI value after paging. On the other hand, both 5G SA implementations always assigned a new and unpredictable 5G-GUTI value. As for the implementation of each operator, the 5G NSA kept the same 5G-GUTI even for 2 days, \textcolor{red}{whereas the 5G SA networks assigned a new value after a time ranging between 90 minutes and 2 hours}, even if neither paging nor a new registration had been made in this period.
Regarding the operator’s deployment scenarios, important differences were observed between the different networks. First, in terms of initial registration, all three 5G SA networks as well as the 5G NSA network, assign a new, unpredictable value of 5G-GUTI. Afterwards, we observed significant differences between the four networks in terms of 5G-GUTI update policy during paging. The 5G NSA and one out of three 5G SA implementations never assigned a new 5G-GUTI value after paging. On the other hand, two 5G SA implementations always assigned a new and unpredictable 5G-GUTI value. As for the implementation of each operator, the 5G NSA and one 5G SA networks kept the same 5G-GUTI even for 2 days, whereas the other two 5G SA networks assigned a new value after a time ranging between 90 minutes and 2 hours, even if neither paging nor a new registration had been made in this period. As for the emulated, OAI-based testbed, it fails to generate a new and unpredictable 5G-GUTI value. For instance, even after a new UE registration, either the previous 5G-GUTI value was used, or a new one almost equal to the previous one.  

\setlength{\leftbarwidth}{5pt}
\setlength{\leftbarsep}{8pt}
\colorlet{leftbarcolor}{blue}

\begin{leftbar}
\noindent \textbf{Takeaway 4:}
The 5G SA and NSA Networks can be 3GPP compliant if properly implementing the 5G-GUTI update mechanism. Only two out of the three tested 5G SA Networks were compliant, whereas the third 5G SA, the NSA and the OAI Networks were found to be weak in terms of implementation of the 5G-GUTI update policy; future work should improve this weakness.
\end{leftbar}

\subsection{Attacks based on lack of ciphering}\label{sec:lack_of_ciphering}

\subsubsection{\textbf{C-RNTI Tracking:}}

%Lack of RRC ciphering is another potential danger in Cellular networks.
C-RNTI is local to the users’ serving BS and is used for the communication between the UE and the RAN as mentioned in Sec.~\ref{Sec:Identifiers}. C-RNTIs can be found in both Uplink 
(UL) and Downlink (DL) RRC messages.
Interestingly,~\cite{jover2016lte} passively analyzed the traffic in LTE and found that the C-RNTI is included without encryption in the header of every single packet, regardless of whether it is signaling or user traffic.
Ultimately, C-RNTI based attacks take advantage of the lack of RRC ciphering, gaining information about the UE location ~\cite{jover2016lte,jover2016lte_first,rupprecht2019breaking,kohls2019lost,rupprecht2020call}. Linkability between the C-RNTI and the victim's phone number or a social network account (e.g.,~Messenger~\cite{MessengerApp}) can be easily done by the adversary with a few messages or calls.
After that, decoding the DL messages including the C-RNTI can be done using passive sniffers that are available~\cite{falkenberg2019falcon,bui2016owl,kumar2014lte}.
C-RNTI attacks lead to location tracking of the victim UE. 
Concluding, the lack of ciphering in RRC  messages is the privacy weakness behind this attack.%~\textcolor{red}{One more Figure showing the difference in ciphering algorithms (Null ciphering against NEA2 ciphering)}

\subsubsection{\textbf{UE Measurement reports:}}

To make matters worse, a more fine-grained attack called UE Measurements reports attack~\cite{olimid2017lowcost} take advantage of the un-ciphered RRC messages, including measurements made by the UE, such as the signal strength of nearby cells, thereby localizing the UE with triangulation.
Compared to the C-RNTI tracking that localizes in a BS level, this attack can estimate the exact coordinates of the UE.
Concluding, the lack of ciphering in RRC messages is the privacy weakness that leads to this attack.

\setlength{\leftbarwidth}{5pt}
\setlength{\leftbarsep}{8pt}
\colorlet{leftbarcolor}{black}

\begin{leftbar}
\noindent
\textbf{5G Enhancement 5: Ciphering of RRC messages}\\
At the gnB level, current 5G Systems are expected to implement New Radio Encryption Algorithm (NEA) 0, 128-NEA1 and 128-NEA2 for confidentiality (ciphering) protection as analyzed in Sec.~5.2.2 of~\cite{3gpp_fake_base_stations}.
Ciphering of RRC messages is optional, so it is up to the operator if it is enabled or not.
\end{leftbar}

\noindent \textbf{Experimental Validation:}
Looking into the RRC Security Mode Command message, we can find the information related to the RRC ciphering.
All setups we examined (i.e., OAI, 5G SA and NSA operator networks) use NEA 0 (Null ciphering) for the ciphering of RRC messages.

\setlength{\leftbarwidth}{5pt}
\setlength{\leftbarsep}{8pt}
\colorlet{leftbarcolor}{blue}

\begin{leftbar}
\noindent \textbf{Takeaway 5:}
Null Ciphering is 3GPP compliant, but we highlight it is not a safe option: the network is vulnerable to C-RNTI tracking and UE measurement reports attacks.
\end{leftbar}

\subsection{Attacks based on lack of integrity}

In this section, we discuss bidding-down attacks~\cite{shaik2015practical,shaik2019new,karakoc2023never} that take advantage of the lack of integrity in messages including the Security and Radio capabilities.
As the name suggests, a bidding-down attack forces the UE to use lower-quality network protocols and mechanisms, resulting, among others, in degradation of user privacy. 

\subsubsection{\textbf{Radio Capabilities attack:}}
In~\cite{shaik2019new,shaik2015practical}, they analyze an attack based on the UE radio capabilities. The adversary, using a rogue base station, intercepts the UE radio capabilities message that is transmitted up to 4G, before the establishment of a secure channel, thus with no integrity protection activated.
The adversary modifies appropriately their characteristics (e.g.,~modify the frequencies supported by the UE Modem), thus, managing to downgrade the UE to use a lower-level cellular network generation.

\setlength{\leftbarwidth}{5pt}
\setlength{\leftbarsep}{8pt}
\colorlet{leftbarcolor}{black}

\begin{leftbar}
\noindent
\textbf{5G Enhancement 6: Transmission of Radio Capabilities over a Secured Channel}\\
The UE radio capabilities were transmitted before the establishment of the RRC security channel, enabling bidding down attacks in 3G and 4G. This mistaken flow was corrected in 5G, as already mentioned in Sec.~\ref{Sec:message_flow}, and depicted in Fig.~\ref{fig:5G_flow}. This is because the RRC UE Capability Inquiry is transmitted after the RRC Security Mode Complete, when integrity protection is enabled. Based on this outlook, the bidding-down attacks based on UE Radio capabilities are eliminated.
\end{leftbar}

\noindent \textbf{Experimental Validation:}
All different real operator networks, and the OAI testbed asked for the Radio Capabilities after the establishment of an RRC secure channel.

\setlength{\leftbarwidth}{5pt}
\setlength{\leftbarsep}{8pt}
\colorlet{leftbarcolor}{blue}

\begin{leftbar}
\noindent \textbf{Takeaway 6:}
The transmission of the Radio Capabilities
after the establishment of an RRC secure channel indicates the elimination of the problem caused in previous generations~\cite{shaik2019new,shaik2015practical}. 
\end{leftbar}

\begin{figure}
\begin{subfigure}{0.9\columnwidth}
  \centering
  \includegraphics[width=.99\linewidth]{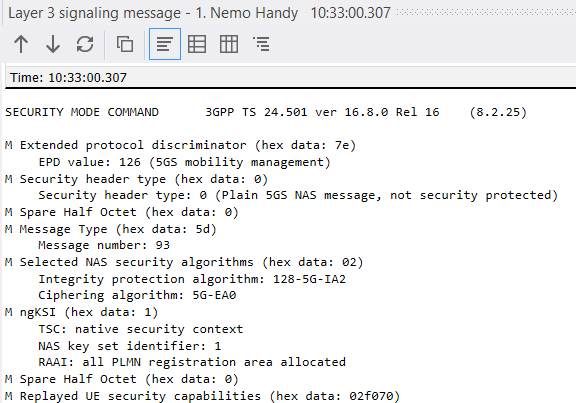}  
  \caption{Enhanced SMC in 5G SA networks. Similar in 5G NSA implementation.}
  \label{fig:smc_enhanced_operator}
\end{subfigure}
\hfill
\begin{subfigure}{0.9\columnwidth}
  \centering
  \includegraphics[width=.99\linewidth]{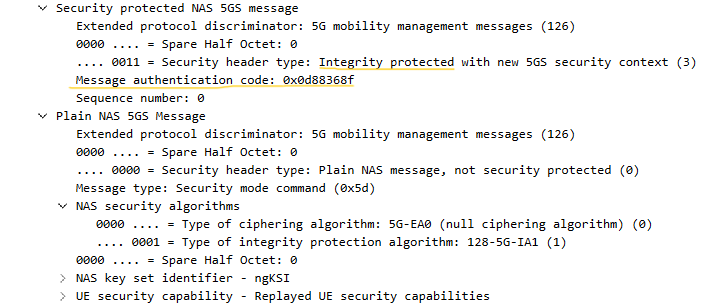}  
  \caption{OAI enhanced SMC.}
  \label{fig:smc_enhanced_OAI}
\end{subfigure}
\hfill
\caption{Enhanced SMC Procedure between operator's implementation and OAI.}
\label{fig:smc_enhanced}
\end{figure}

\subsubsection{\textbf{Security Capabilities attack:}}\label{Sec:Security_Capabilities_attack}

To begin with, the adversaries described in ~\cite{shaik2019new,shaik2015practical} exploit the security capabilities in the first registration message.
The fact that the integrity protection of NAS messages is mandatory is not enough for the initial NAS message protection, since at this stage the UE and the network have not yet defined the algorithms that will be used for integrity and ciphering.
An active adversary with a rogue base station can intercept the Registration Request message, modify the UE Security Capabilities (e.g.,~disable them) and release the modified message to the legitimate network.
As a result the UE obtains weaker security compared to the one that their phone can support. 
%No privacy mechanism exists for detecting this attack, so the UE obtains weaker security compared to the one that their phone can support. 

\setlength{\leftbarwidth}{5pt}
\setlength{\leftbarsep}{8pt}
\colorlet{leftbarcolor}{black}

\begin{leftbar}
\noindent
\textbf{5G Enhancement 7: Enhanced SMC Privacy}\\
A new, enhanced and integrity-protected initial NAS message transmission procedure has been introduced in the Sec.~6.7.2 of~\cite{3gpp_fake_base_stations} to defend the bidding-down attacks, described in~\cite{karakoc2023never,shaik2015practical,shaik2019new}. This procedure applies in the NAS Security Mode Command (SMC) message.
In fact, the CN initiates the integrity protection and transmits among others a Message Authentication Code (MAC), the UE replayed the Security Capabilities that were received in the initial Registration Request message by the UE, and the security algorithms that will be used for integrity and ciphering. The UE uses the MAC to verify the integrity of the NAS SMC message and if the integrity verification is successful.
Then, it verifies that the Security Capabilities replayed by the CN are the ones that had been originally transmitted in the Registration Request message. If everything is correct, the UE answers with a NAS Security Mode Complete and the establishment of a secure NAS channel is completed. 
After that, any attempt of the adversary to modify any NAS message is futile, since NAS integrity is mandatory, as mentioned earlier.
\end{leftbar}

\noindent \textbf{Experimental Validation:}
We verified the NAS Security mode command in all of our networks. We observed that all replay the initially transmitted security capabilities but only OAI follows the procedure described in the Sec.~6.7.2 of ~\cite{3gpp_fake_base_stations} correctly.
As shown in Fig.~\ref{fig:smc_enhanced_operator}, all operator's networks omit the MAC code integrity protections and simply replay the UE network capabilities received in the registration message and mention the algorithms used for integrity and ciphering.
On the other hand, Fig.~\ref{fig:smc_enhanced_OAI} shows that OAI follows the enhanced SMC Privacy procedure accurately, transmitting the replayed UE Capabilities, and the chosen security algorithms along with the MAC code that ensures the integrity protection.

\setlength{\leftbarwidth}{5pt}
\setlength{\leftbarsep}{8pt}
\colorlet{leftbarcolor}{blue}

\begin{leftbar}
\noindent \textbf{Takeaway 7:}
All of the implemented solutions defend the Security Capabilities bidding-down attack as mentioned in~\cite{shaik2015practical,shaik2019new}, but all operator's implementations (NSA and SA) are vulnerable to another, novel attack, as will be explained next in Sec.~\ref{Sec:attack_bidding_down}.
\end{leftbar}

\section{New Vulnerabilities of 5G}
\label{sec:new_vulnerabilities}

In this section, we analyze two new vulnerabilities based on the measurement studies described earlier.

\subsection{GUTI Reallocation Command Attack}\label{Sec:GUTI_Reallocation_attack}

\begin{figure}
  \centering
  \includegraphics[width=0.9\linewidth]{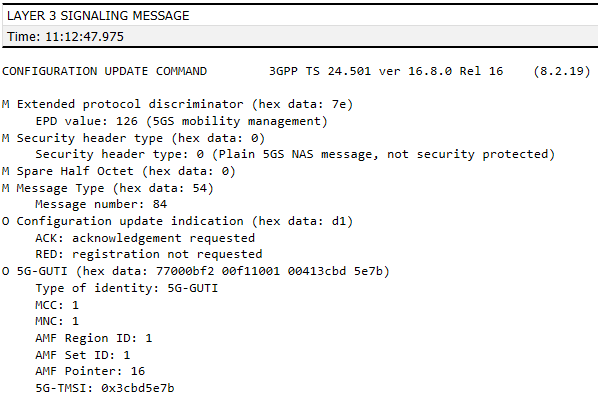} 
  \caption{5G-GUTI Reallocation Command in  5G SA operator's networks.}
  \label{fig:GUTI_problem}
  \vspace{-5mm}
\end{figure}

As shown earlier, two out of three 5G SA networks always assign a new,
unpredictable 5G-GUTI value to the UE. This new value is transmitted to the UE through a NAS Configuration Update Command message. Unfortunately, these two 5G SA networks transmit this command without security protection, neither in terms of integrity nor of ciphering, to the UE causing a significant privacy risk.
Fig.~\ref{fig:GUTI_problem} illustrates the Nemo Handy capture that demonstrates the aforementioned privacy vulnerability.
We highlight that Arapinis et al.~\cite{arapinis2014privacy} had referred
to this danger in previous generations, but later the same authors characterized this attack as a mere theoretical one \cite{arapinis2017analysis}, since
the previous generations were not assigning new 5G-GUTI
values.
As we saw in our experiments, 5G SA assigns new 5G-GUTI values.
Thus, we demonstrate this vulnerability in a real scenario for the first time. 

A potential adversary can take advantage of two vulnerabilities leading to two different types of attacks. First, the lack of integrity means that the content of the above mentioned message can be modified. Therefore, a Man-in-the-Middle (MiTM) attacker can intercept a Configuration Update Command message and modify the 5G-GUTI value to a different one. Then, when the UE tries to reestablish some content with the network, using this modified 5G-GUTI value, the network will not be able to recognize him/her, thus leading to a Denial of Service (DoS) attack. On the other hand, we stress that an attack called GUTI Refreshment Neutralization that was presented in 5G networks as a proof of concept in~\cite{hussain20195greasoner}, exploits the Configuration Update Command as well. In that work, the authors took advantage of the lack of Acknowledgment (ACK) request in the Configuration Update Command. The lack of this parameter means that the UE does not send a Configuration Complete message as an answer to the Configuration Update Command. As a result, the adversary was capable of intercepting or dropping the Configuration Update Command without having to send an Acknowledgment Request to the CN. The UE was never obtaining the Configuration Update Commands, and as a result, its GUTI remained the same. As we see in Fig.~\ref{fig:GUTI_problem}, an ACK is requested, so the GUTI Refreshment Neutralization attack is not applicable as opposed to our attack that exploits the lack of integrity in the Configuration Update Command message.

Furthermore, the adversary can take advantage of the lack of ciphering
in the 5G-GUTI Reallocation Command, aiming to track the victim’s location. First, the adversary can send silent messages or perform silent calls to the victim. A silent call or message is one that does not trigger any notification to the recipient UE~\cite{hong2018guti}. After making (sending) these silent calls (messages), the
paging mechanism is triggered either in Cell or TA level, followed by the unencrypted Configuration Update Command including the new victim’s 5G-GUTI value. The adversary can sniff the downlink paging
channel and search for the victim’s unencrypted 5G-GUTI value, thus verifying the victim’s presence in a specific cell or TA. The realization of this attack in a large scale network, where many devices are connected, is difficult. 
By the time the adversary sends the message to the PoI, other devices in the same area can receive messages as well, so the adversary may see many different Configuration Update Commands. 
To enhance the attack's accuracy, the attacker can send messages with a specific frequency (timing attack) aiming to estimate the victim's paging time, or during non-busy time windows in terms of traffic (e.g.,~very late at night), or by avoiding to search for the victim in very crowded areas. Unfortunately, with the current experimental setting it was impossible to verify the accuracy of this attack in the private operator deployments, since the traffic from UEs was minimal.

Based on related 3GPP documents~\cite{3gpp_fake_base_stations, 3GPP_NAS}, the responsibility behind these attacks can be placed to two factors.
First, the operator \textbf{must not} transmit NAS messages without integrity.
In fact, as shown in Fig.~\ref{fig:smc_enhanced_operator}, during the establishment of the NAS Security Channel, integrity was used. 
Unfortunately, when the UE switched from "Idle" to "Connected" state, due to the activation of the paging mechanism, the operator lost the NAS Security context and did not reestablish it with a NAS SMC message.
Second, the lack of ciphering comes from an omission in 3GPP documents.
As explicitly written in Sec.~5.5.1 of~\cite{3gpp_fake_base_stations}: "NAS confidentiality is optional to use", also similarly stated in Sec.~4.4.1 of~\cite{3GPP_NAS}: "The use of ciphering in a network is an operator option. Operation of a network without ciphering is achieved by configuring the AMF so that it always selects the "null ciphering algorithm", 5G-EA0".

\setlength{\leftbarwidth}{5pt}
\setlength{\leftbarsep}{8pt}
\colorlet{leftbarcolor}{blue}

\begin{leftbar}
\noindent \textbf{Takeaway 8:}
This is a potential future threat to address: it is not the best strategy to transmit the Configuration Update Command, including the reallocated value of the privacy-sensitive identifier 5G-GUTI, without integrity and encryption.
\end{leftbar}

\begin{figure*}[t]
\begin{subfigure}[b]{0.99\linewidth}
  %\centering
  \raggedleft
  \includegraphics[width=0.9\textwidth]{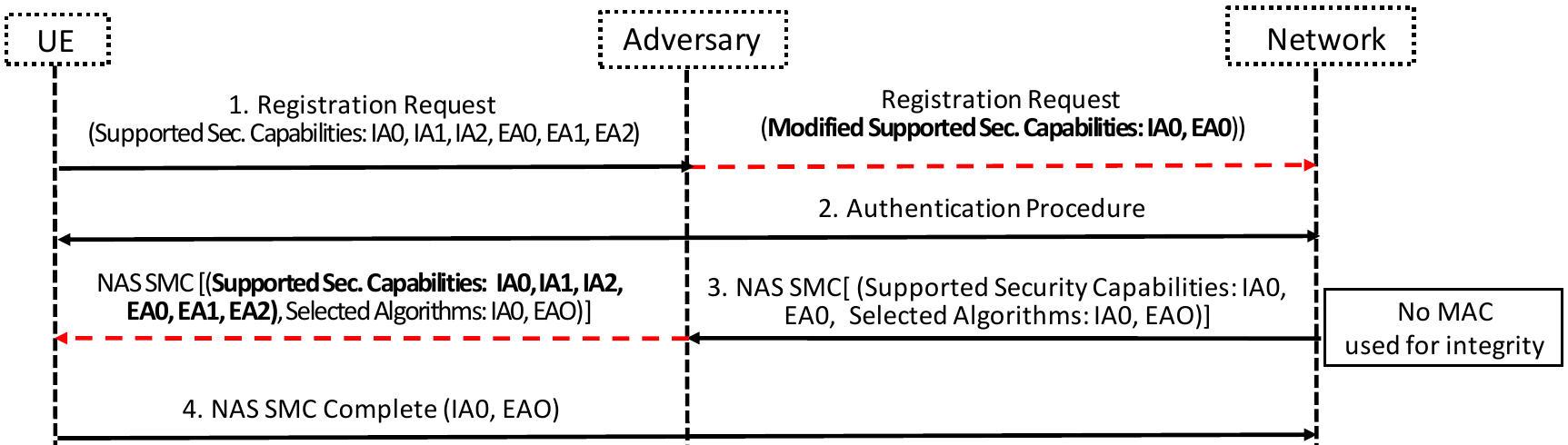}  
  \caption{Successful Security Capabilities Bidding-Down attack in operator's 5G SA and NSA networks.}
  \label{fig:operator_bidding_down_success}
\end{subfigure}
\hfill
\begin{subfigure}[b]{0.99\linewidth}
  \centering
  \includegraphics[width=1\textwidth]{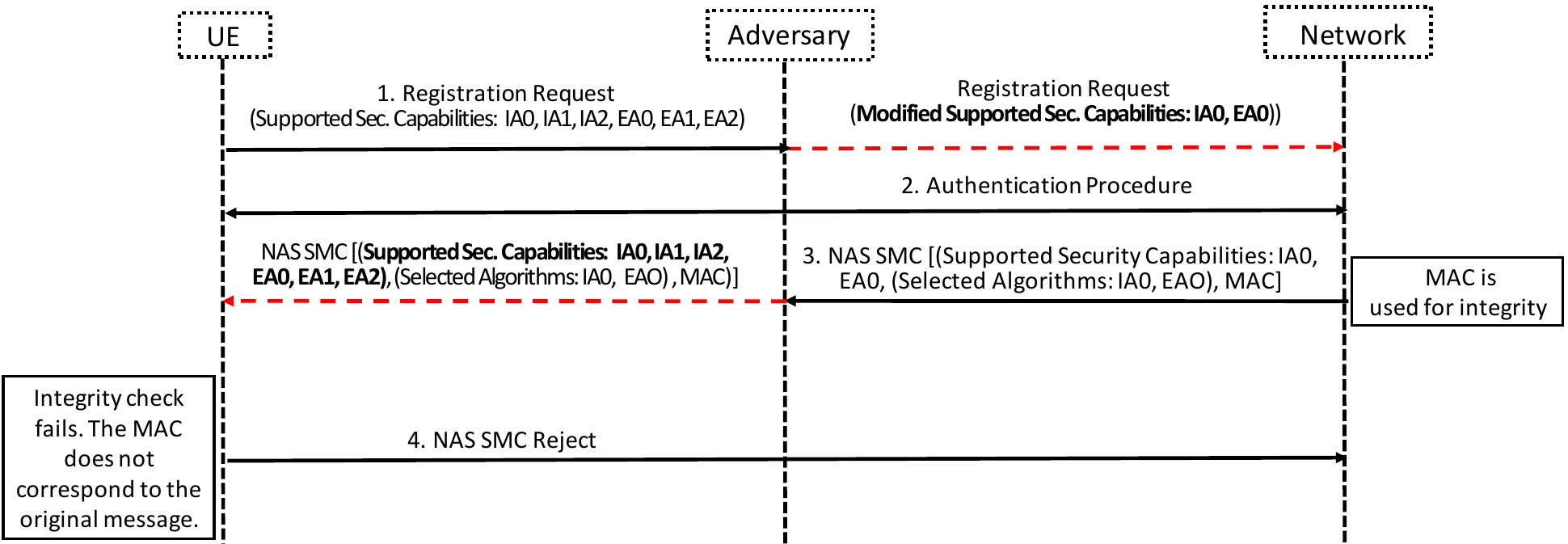}  
  \caption{Unsuccessful Security Capabilities Bidding-Down attack in OAI emulated network.}
  \label{fig:OAI_bidding_down_fail}
\end{subfigure}
\hfill
\caption{The importance of the MAC code in the NAS SMC Enhanced Procedure.}
\vspace{-3mm}
\label{fig:New_Bidding-Down_attack} 
\end{figure*}

%\vspace{-3mm}
\subsection{Security Capabilities Bidding-Down Attack}\label{Sec:attack_bidding_down}

As it was shown in Sec.~\ref{Sec:Security_Capabilities_attack}, the enhanced NAS Security Mode Command procedure followed  by all different operator's implementations (5G NSA and three 5G SA networks), and shown in Fig.~\ref{fig:smc_enhanced_operator}, is weaker compared to the one presented in the 3GPP TS. 33.501~\cite{3gpp_fake_base_stations}. 
However, the same procedure is implemented correctly in OAI, as shown in Fig.~\ref{fig:smc_enhanced_OAI}.
Here, we discuss a proof of concept of a Bidding-Down attack that works as an extension to the initially presented ones~\cite{shaik2015practical,shaik2019new}, and can lead to privacy deterioration.

First, as shown in both Figures~\ref{fig:operator_bidding_down_success} and ~\ref{fig:OAI_bidding_down_fail}, the UE sends a Registration Request indicating the supported Security Capabilities for ciphering (encryption) algorithms (e.g., EAO, EA1, EA2) and integrity algorithms (IAO, IA1, IA2) to the network.
An active MiTM adversary intercepts this message similarly to~\cite{shaik2015practical,shaik2019new} and modifies the supported security algorithms using the weakest ones, in this case IA0 with EA0.
Afterwards, an authentication procedure based on 5G-AKA is followed as described in the Sec.~\ref{Sec:message_flow}. 
When the authentication is completed, the CN sends a NAS Security Mode Command (SMC) to the UE. The 5G NSA and SA operator's networks include only the replayed security capabilities and the selected security algorithms, as illustrated in Fig.~\ref{fig:operator_bidding_down_success}.
As explained, the OAI implementation follows exactly the 3GPP related documentation transmitting the NAS SMC message with its corresponding Message Authentication Code (MAC) for integrity protection.

Afterwards, the adversary modifies again the supported capabilities, as we see in both Figures~\ref{fig:operator_bidding_down_success} and~\ref{fig:OAI_bidding_down_fail}, so that the UE cannot understand the difference between the Supported Security Capabilities as transmitted in the first Registration message and the ones received in the NAS SMC message. In all different 5G SA and NSA operator's implementations, there is not a MAC code, so the integrity of the NAS SMC message cannot be verified by the UE.
As a result, the UE accepts the NAS SMC procedure and the weakest security algorithms are used for integrity and ciphering, leading to a successful biding-down attack.
Given that the OAI implementation includes the MAC code in the process, the UE can verify the integrity of the NAS SMC message though the MAC code using its private key. The adversary cannot have access to this key, so it cannot modify the MAC code.
As a result, the UE understands the NAS SMC message modification by the adversary, the integrity check fails and the UE rejects the NAS SMC.

\setlength{\leftbarwidth}{5pt}
\setlength{\leftbarsep}{8pt}
\colorlet{leftbarcolor}{blue}

\begin{leftbar}
\noindent \textbf{Takeaway 9:}
An operator CN implementation may omit the transmission of the Message Authentication Code (MAC) in the NAS Security Mode Command, potentially making both vulnerable to a Security Capabilities-based bidding-down attack.
The OAI implementation is adjusted to 3GPP specifications and can mitigate this attack.
\end{leftbar}

\section{Related Work}\label{sec:rel-work}

Next, we cover related studies grouped into four main areas, and outline how they compare to our work.

\noindent
\textbf{Measurement Studies:} 
There are some experimental studies in 5G NSA and SA  networks, mentioning the privacy vulnerabilities of these network infrastructures.
For instance,~\cite{lasierra2023european,wani2024security,park20215g,palama2021imsi} focused only on 5G NSA networks' security vulnerabilities, whereas ~\cite{nie2022measuring} focused only on 5G SA networks' ones.
In contrast to them, we compare 5G SA and NSA operator networks under the same operating settings, analyzing a wide range of security features and their corresponding privacy attacks, thus providing a more concrete and complete work on this topic than the previously mentioned works. We also include in our study a 5G SA OAI open source framework setup, showing the degree of its compliance with the 3GPP standards, and thus its utility to the research community studying these issues.

\noindent
\textbf{Attacks on permanent UE IDs:} Many different adversaries have tried to steal the UE IMSI, either through an IMSI catching attack~\cite{strobel2007imsi,meyer2004man,kotuliak2022ltrack,erni2022adaptover,hussain2019privacy,palama2021imsi}, or via an IMSI paging attack~\cite{kune2012location,arapinis2017analysis,shaik2015practical,bojic2017opportunities}.
Our work shows that IMSI catching attack, as described in Sec.~\ref{Sec:IMSI_Catchers}, exists only in 5G NSA networks, since in 5G SA networks SUCI security characteristic is implemented.
In fact, \textit{we are the first to show that SUCI can be supported by a real operator 5G SA CN}.
As for the IMSI paging attack, as described in Sec.~\ref{Sec:IMSI_paging_attack} it is mitigated in both 5G SA and NSA networks, since IMSI was never used for paging.
Also, IMEI catching was of primary interest to previous generations' adversaries~\cite{dabrowski2014imsi,olimid2017lowcost,michau2016not,park2022doltest}, but as shown in our work in Sec.~\ref{IMEI_Catching}, both 5G SA and NSA networks have eliminated the vulnerabilities that led to this attack.

\noindent
\textbf{Attacks based on TMSI/GUTI and lack of ciphering:}
Previous cellular network generations focused on TMSI-related attacks~\cite{kune2012location,arapinis2012new,arapinis2014privacy,hong2018guti}, taking advantage of the weak TMSI reallocation mechanism. As analyzed in Sec.~\ref{Sec:TMSI_attack}, these attacks cannot be deployed in a 5G SA network that follows the 3GPP standards for the reallocation of 5G-GUTI.
Also, C-RNTI tracking~\cite{jover2016lte_first,jover2016lte,rupprecht2019breaking,rupprecht2020call,ludant20235g} and the UE measurements reports attacks~\cite{shaik2015practical,olimid2017lowcost,bitsikas2021don} took advantage of the lack of ciphering, as analyzed in Sec.~\ref{sec:lack_of_ciphering}. Our work shows that this kind of problems continue to exist in 5G networks, and thus, 3GPP should think of mandating ciphering.
Finally, a recent work in 5G~\cite{hussain20195greasoner} theoretically showed potential problems with the GUTI reallocation procedure due to the lack of ACK request. As shown in Sec.~\ref{Sec:GUTI_Reallocation_attack}, even if the ACK request is included in the Configuration Update Command, another privacy vulnerability can persist in the GUTI reallocation process, due to lack of integrity and ciphering. %In fact, as explained in~\ref{Sec:GUTI_Reallocation_attack}, this vulnerability can lead either to DoS or to UE tracking attacks.

\noindent
\textbf{Bidding-Down Attacks:} There are a few works focusing on bidding-down attacks based on Security Capabilities as presented in Sec.~4.4.2.
First,~\cite{shaik2015practical} introduced this attack as in LTE networks, proposing the replay of the Security capabilities for its mitigation. Then,~\cite{chlosta2019lte} and~\cite{karakoc2023never} showed the lack of a replay mechanism for security capabilities in 4G, and the lack of enhanced initial NAS message protection in 5G NSA networks, respectively.
Our work shows the replay of security capabilities has been done in both 5G NSA and SA, but as explained in Sec.~\ref{Sec:attack_bidding_down}, it is not enough.
In fact, we are the first to stress the importance of Message Authentication Code (MAC) process in the NAS SMC command, and show its correct implementation in an OAI emulated network.

\vspace{-3mm}
\section{Discussion}\label{sec:discussion}

%In this section, we include the lessons learned from our work, elaborating on the key takeaways presented in earlier Sections~\ref{Sec:pre-5g-attacks} and~\ref{sec:new_vulnerabilities}. %First, we include our afterthought about the privacy level of 5G SA and NSA networks. We also analyze the perfomrance of OAI the de facto open source software for 5G in terms of privacy. Then, we highlight important 3GPP omissions in the related 5G TSs.

Based on the qualitative and experimental analysis presented in Sec.~\ref{Sec:pre-5g-attacks}, and the corresponding summary made in Table~\ref{tab:Overview}, we draw important lessons about 5G privacy.
First, all three deployment setups (operator's 5G SA and 5G NSA networks, and OAI 5G SA SDN network) have mitigated traditional attacks caused by the IMSI paging and IMEI or Radio Capabilities transmission, following the corresponding mitigation mechanisms, as mentioned in Takeaways $2$, $3$ and $6$.
Second, the 5G SA operator's implementations and the OAI 5G SA SDN testbed support SUCI mechanism (Takeaway $1$), thus, mitigating IMSI Catchers attack and offering better UE identity privacy, as opposed to the 5G NSA infrastructure.
Third, the 5G-GUTI update mechanism is implemented accurately in two out of three 5G SA operator's networks, thus, mitigating the TMSI Deanonymity attack, in contrast to the 5G NSA operator network and OAI SDN based one (Takeaway $4$).
Unfortunately, C-RNTI tracking and UE measurement reports attacks are still existent in all of the examined networks.
Therefore, stricter 3GPP regulations are needed, mandating the ciphering of RRC messages (Takeaway $5$).
Finally, NAS SMC procedure is partially implemented by all 5G SA networks and the 5G NSA network as well, whereas OAI followed exactly the 3GPP TSs., adding the MAC in corresponding messages (Takeaway $7$).

Our work also presented two new 5G vulnerabilities that can be potential privacy risks in future 5G networks.
As explained in Sec.~\ref{Sec:GUTI_Reallocation_attack} and summarized in Takeaway $8$, 5G SA networks transmit the Configuration Update Command without integrity and ciphering, thus, risking DoS attacks and potential UE tracking.
We believe that measures should be taken against this problem, both from the operators (integrity protection) and 3GPP community (mandatory ciphering) sides.
Moreover, in Sec.~\ref{Sec:attack_bidding_down}, another novel privacy attack was outlined, taking advantage of the NAS SMC vulnerability of all operator's networks.
As shown, including the MAC in the NAS SMC message, similarly to what OAI implementation does, is of paramount importance for being safe against bidding-down attacks (Takeaway $9$).

Next, we discuss why these privacy problems arose. Regarding the use of RRC ciphering, as explained in 3GPP-related documents~\cite{3gpp_fake_base_stations} and discussed earlier (Sec.~\ref{Sec:pre-5g-attacks}), it is a functionality that remains optional for the operator. 
This can explain its lack of activation during our experiments, even if it is supported by the gNBs. This finding aligns well with earlier reports~\cite{lasierra2023european,chlosta2019lte},
that showed operators are often unwilling to activate such optional security features that add computational overhead to the network. 
As for the attack described in Sec.~5.1, the different CNs lost the integrity protection that had been activated before with the NAS Security Command message (Fig.~\ref{fig:smc_enhanced_operator}). 
We confirmed this was a result of a CN mis-configuration. 
Further, the lack of NAS ciphering is an operator's choice (Sec. 5.1), and similar to RRC ciphering, it was not activated by the different CNs for similar reasons. 
Finally, the 5G bidding-down attack, described in Sec.~\ref{Sec:attack_bidding_down}, exploits another CN mis-implementation, which is the lack of MAC in the NAS Security Mode Command message. 
We remark that the 3GPP-related document~\cite{3gpp_fake_base_stations} does not sufficiently explain the importance and use of this MAC in this process, and therefore, the operators/vendors may under-estimate its critical role. 
As shown in Sec.~\ref{sec:new_vulnerabilities}, both new attacks were applicable to different 5G networks, indicating common CN mis-configurations/mis-implementations for the specific operator and vendors. 
We stress the importance of our findings: same vendors of 5G CN solutions can be providers to network operators in various countries, and therefore, such privacy attacks can be applicable to other CNs beyond the ones tested.

%\balance 
To summarize, 5G SA followed almost all of the 3GPP privacy enhancements mentioned in Sec.~\ref{Sec:pre-5g-attacks} and is more privacy-preserving than the 5G NSA network, which inherently missed SUCI and 5G-GUTI update mechanisms.
Furthermore, stricter policies are needed from the 3GPP community, mandating ciphering and mitigating the corresponding privacy attacks. Operators and 3GPP community should take into consideration the attacks mentioned in Sec.~\ref{sec:new_vulnerabilities} and make the appropriate adjustments to their implementations and Technical Specifications, respectively. 
Finally, OAI showed almost excellent results in terms of privacy, considering that it is an open source project, missing only the 5G-GUTI update mechanism and can be used for privacy studies by the research and industry 5G community, while the considerations from the present study are taken into account.

We mention that all of our findings in the private operator’s 5G NSA and
SA networks have been reported to the operator for further assessment. 
Besides, the weakness of 5G-GUTI update mechanism has been communicated to OAI development team as well, together with details on the procedure to correct it.

\textbf{AKA Protocol Linkability attack:}
We have not discussed here the AKA protocol linkability attack~\cite{arapinis2012new,arapinis2017analysis,khan2014another,hahn2014privacy}.
This attack was common since 3G (no AKA protocol in 2G), performing location tracking by linking two AKA sessions of the same user.  Recent papers~\cite{fei2023vulnerability,wang2021privacy,hussain20195greasoner,koutsos20195g} verified it in 5G.
The 3GPP community analyzed this attack in Sec.~6.2 of~\cite{suci_related_solutions}, proposing modifications to the 5G AKA protocol, such as ciphering of different 5G-AKA parameters.
These modifications can mitigate this problem, but are still not included in the 5G Systems Security protocols and procedures in 3GPP 5G Specifications~\cite{3gpp_fake_base_stations}.
Thus, until now, there is no official 3GPP mitigation mechanism against this attack.
%\vspace{4mm}
%On the other hand, as we see in the private operators' network, some gaps still exist. Stricter 3GPP regulations and pressure to the operators to be careful with their implementations are needed.

%OAI follows a lot of privacy characteristics, except of ..... It can be used for privacy studies and we plan to improve their existing weaknesses as a future work with the appropriate code modifications.

\vspace{3mm}
\section{Conclusion}

In this paper, we presented a first of its kind, head-on comparison between 5G stand alone (SA) and non-stand alone (NSA) cellular networks, in terms of security and privacy capabilities across 8 different top pre-5G attacks.
As shown, 5G SA offers higher identity privacy, since SUCI identifier and 5G-GUTI reallocation mechanism are properly supported.
On the other hand, problems inherited by the previous cellular generations, and caused by the lack of ciphering continue to exist, since ciphering is still optional in 3GPP documentation, making both of the 5G implementations (SA and NSA) vulnerable to NAS and RRC based protocol attacks.
Our in-depth analysis also revealed two new potential attacks against UE privacy.
The first exploits the lack of ciphering and integrity of the Configuration Update Command used for updating the 5G-GUTI value.
The second is a security capabilities bidding-down attack, exploiting the lack of inclusion of Message Authentication Code (MAC) in the NAS Security Mode Command.
Finally, privacy features of OAI were analyzed showing that it can be considered for privacy studies, as only the 5G-GUTI update mechanism is missing.

\vspace{3mm}
\section*{ACKNOWLEDGEMENTS}
This research was partially supported by:
The Ministry of Economic Affairs and Digital Transformation of Spain and the European Union-Next Generation EU programme for the Recovery, Transformation and Resilience Plan and the Recovery and Resilience Mechanism under agreements TSI-063000-2021-63 (MAP-6G), TSI-063000-2021-142 and TSI-063000-2021-147 (6G-RIEMANN);
The European Union Horizon 2020 program under grant agreements 101021808 (SPATIAL), 101096435 (CONFIDENTIAL6G) and 101139067 (ELASTIC).
The views and opinions expressed are those of the authors only and do not necessarily reflect those of the European Union. Neither the European Union nor the granting authority can be held responsible for them.
%
%The research conducted by IMDEA Networks was sponsored in part by the project MAP-6G, reference TSI-063000-2021-63, granted by the Ministry of Digital Transformation and Public Service, and the European Union-NextGenerationEU through the UNICO-5G R\textbackslash\&D program of the Spanish Recovery, Transformation and Resilience Plan.
%
The authors thank Maria del Mar Moreno, Javier Jimemez, Manuel Palacios Trapero for their help during the experiments. 

\newpage
\bibliographystyle{ACM-Reference-Format}
\bibliography{main.bib}

\end{document}